\documentclass[
    aps,
    prb,
    twocolumn,
    10pt,
    longbibliography,
    floatfix,
    nofootinbib,
]{revtex4-2}

\usepackage[usenames, dvipsnames]{color}
\usepackage{physics}
\usepackage[english]{babel}
\usepackage[utf8]{inputenc}
\usepackage{amssymb}
\usepackage{amsmath}
\usepackage[pdftex]{graphicx}
\usepackage{xspace}
\usepackage{bm}
\usepackage{hyperref}
\usepackage{soul}
\hypersetup{
    colorlinks,%
    citecolor=blue,%
    filecolor=blue,%
    linkcolor=blue,%
    urlcolor=blue
}

\graphicspath{{.}{Figs/}}

\begin{document}

\title{Jittery Quantum Boomerang Effect}

\author{Pedro Dornelas}
\affiliation{Instituto de F\'{i}sica, Universidade Federal de Uberl\^{a}ndia, Uberl\^{a}ndia, Minas Gerais 38400-902, Brazil}

\author{Gerson J. Ferreira}
\affiliation{Instituto de F\'{i}sica, Universidade Federal de Uberl\^{a}ndia, Uberl\^{a}ndia, Minas Gerais 38400-902, Brazil}

\date{\today}
\begin{abstract}
We study the dynamics of a spin-polarized wave packet in a disordered Rashba two-dimensional electron gas and identify a \textit{jittery quantum boomerang effect} in which longitudinal and transverse motion return to the origin through fundamentally distinct mechanisms. Starting from an initial state with finite momentum along $x$ and spin polarized along $z$, we calculate the time evolution by combining a Chebyshev expansion of the time-evolution operator with a disorder ensemble average. In the weak-scattering regime, equations of motion derived from the quantum kinetic equation reproduce the numerical trends and show that impurity scattering acts as a viscous damping mechanism that suppresses the transient Zitterbewegung and drives the transverse displacement back to $y=0$ at long times. In contrast, the longitudinal dynamics show a Drude-like saturation at weak disorder. These results are consistent with the vanishing intrinsic spin Hall conductivity in the disordered Rashba model and with experimental observations of a transient intrinsic spin Hall effect in the time-domain. As disorder increases, the longitudinal dynamics evolve to a partial return toward the origin, which signals a transition from weak antilocalization to Anderson localization in 2D.
\end{abstract}

\maketitle

\section{Introduction}

Anderson localization (AL) is a central paradigm of quantum transport, in which interference suppresses diffusion in disordered media \cite{anderson1958absence}. It is a universal wave phenomenon that spans a wide range of physical platforms, including electrons, photons, ultrasound, and ultracold atoms \cite{wiersma1997localization, weaver1990anderson,billy2008direct, roati2008anderson}. The seminal 1979 \textit{Gang of Four} work
\cite{abrahams1979scaling} established the scaling theory of localization in one, two, and three dimensions. In two dimensions (2D), this theory predicts marginal localization: all states are localized for any finite disorder, but with a localization length that can be exponentially large in the weak-disorder regime. In the presence of strong spin-orbit coupling (SOC), however, spin rotation introduces a $\pi$ phase shift that reverses the sign of the quantum-interference correction to conductivity \cite{hikami1980spinorbit}, leading to weak antilocalization (WAL) in Rashba 2D electron gases (2DEGs) \cite{marinescu2019closedform}, graphene \cite{tikhonenko2008weak,yan2023weak, golub2024theory}, and topological insulators \cite{lu2014weak, dequeiroz2026weak,nag2026weak} and superconductors \cite{wang2022observation, Al_Tawhid_2022,lin2023twodimensionallikesuperconductingproperties}. 
This SOC-disorder interplay is also central in spintronics. 
For instance, the Datta-Das spin transistor \cite{datta1990electronic} can be made robust against disorder in the persistent spin helix regime \cite{schliemann2003nonballistic, bernevig2006exact, koralek2009emergence, weber2007nondiffusive, walser2012direct, ishihara2013direct, fu2016persistent, altmann2016currentcontrolled, dettwiler2017stretchable, ferreira2017spin, weigele2020symmetry, deassis2021spin}.
In topological Anderson insulators \cite{li2009topological, groth2009theory, assuncao2024phase, lopes2026engineering2}, disorder renormalizes the Hamiltonian and induces topological phase transitions.
In the spin Hall effect (SHE) \cite{nagaosa2010anomalous, sinova2015spin}, disorder and SOC jointly determine the intrinsic and extrinsic contributions to the spin Hall conductivity. In particular, for a disordered Rashba 2DEGs, the intrinsic contribution is exactly canceled by vertex corrections due to disorder \cite{mishchenko2004spin, rashba2004sum, inoue2004suppression, raimondi2005spinhall, raimondi2006quasiclassical, engel2007theory, yang2008intrinsic}. 
However, in the time domain, a transient transversal displacement of the electron packet has been observed and attributed to the intrinsic SHE \cite{werake2011observation}. This transversal motion resembles the ballistic side jump \cite{schliemann2007ballistic} caused by the Zitterbewegung oscillations in clean systems \cite{bernardes2006spin, zawadzki2011zitterbewegung, vignale2016theory, ferreira2018zitterbewegung, lavor2021zitterbewegung, hu2025zitterbewegung, predin2026chiralityzitterbewegungrelationberry, culcer2026zitterbewegungvelocitysemiclassicalelectron}.


Recently, a novel dynamical signature of Anderson localization was identified in spinless systems: the \textit{quantum boomerang effect} (QBE) \cite{prat2019quantum}, in which a wave packet launched with finite momentum first drifts and then makes a U-turn to return to the origin. The QBE has been observed in cold-atom and photonic platforms \cite{sajjad2022observation, hou2026quantum}. Other recent theoretical works have discussed QBE robustness under time-reversal-symmetry breaking \cite{janarek2022quantum, noronha2022ubiquity}, interacting systems \cite{janarek2020quantum,janarek2023manybody}, 1D spinful dynamics \cite{janarek2023berezinskii, capuzzi2024spincharge}, classical stochastic and quantum walk frameworks \cite{zamora2025boomerang,buarque2025boomerang}, non-Hermitian systems \cite{noronha2022robust}, and different models of disorder \cite{tessieri2021quantum}.

In this paper we investigate the interplay between Anderson localization and spinful wave-packet dynamics in a disordered Rashba 2DEG. We identify what we term the \textit{jittery quantum boomerang effect}. In this regime, we show that an initial wave packet with finite momentum along the $x$ direction and spin polarized along $z$ exhibits a boomerang-like return to the origin, but with distinct mechanisms governing the longitudinal ($x$) and transverse ($y$) motion. The longitudinal dynamics show a partial return to the origin, i.e., a marginal quantum boomerang effect.
In contrast, the transverse motion displays a transient Zitterbewegung that is suppressed by impurity scattering at long times, driving the center of mass back to $y=0$ and quenching the ballistic side jump \cite{schliemann2007ballistic}. This real-space return to the origin provides a dynamical manifestation of the vanishing intrinsic spin Hall conductivity in disordered Rashba systems 
\cite{inoue2004suppression, nagaosa2010anomalous, sinova2015spin, mishchenko2004spin, rashba2004sum, raimondi2005spinhall, raimondi2006quasiclassical, engel2007theory, yang2008intrinsic} 
and aligns with experimental reports of transient intrinsic spin Hall signals in the time domain \cite{werake2011observation}. To obtain these results, we combine large-scale numerical simulations using a Chebyshev polynomial expansion of the time-evolution operator \cite{tal-ezer1984accurate, fehske2009numerical}, an analytical framework based on the quantum kinetic equation \cite{rammer1986quantum, rammer2007quantum, haug2008quantum, shen2014theory}, symmetry analysis, and localization arguments \cite{prat2019quantum}. This approach reveals that while the longitudinal return is driven by Anderson localization, the transverse return to the origin is enforced by symmetry constraints on the equations of motion, which act as a viscous damping of the Zitterbewegung oscillations.

\section{Numerical model}

\begin{figure}[tb]
    \centering
    \includegraphics[width=\columnwidth]{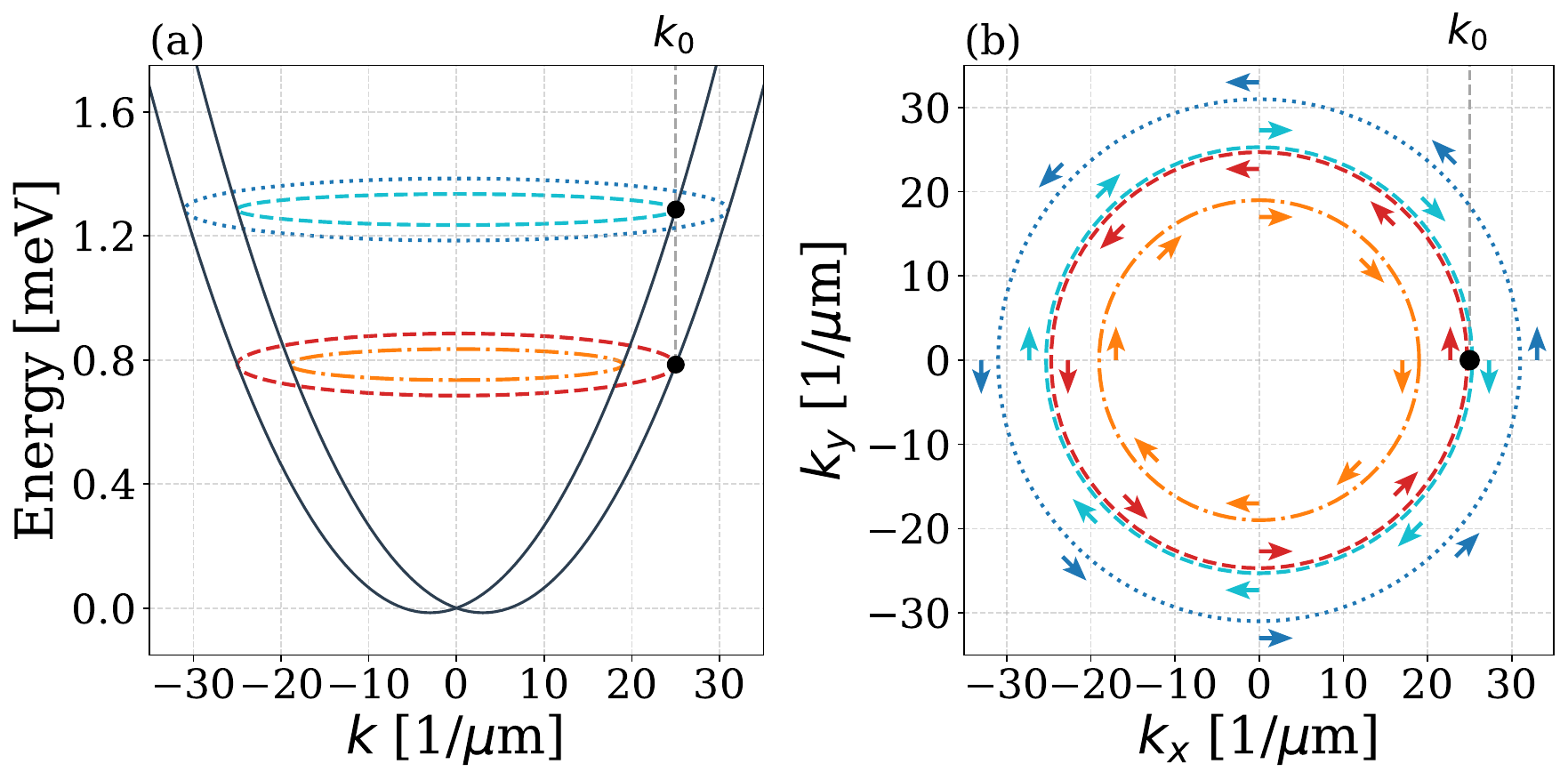}
    \caption{(a) Band structure of the clean Rashba 2DEG, showing the spin-orbit-split bands. The black dots illustrate the initial condition $\ket{\bm{k}_0, \uparrow} = (\ket{\bm{k}_0, +} + \ket{\bm{k}_0, -})/\sqrt{2}$ as a combination of opposite-helicity ($\pm$) states that forms an initial spin-up state ($\uparrow$). For weak scattering, the wave-packet dynamics remain concentrated around the constant-energy circles (dashed colored lines), whose spin texture in $k$ space is shown in (b). The central (red and light blue) circles have the same radius $k_0$, but are slightly split in the figure for clarity.}
    \label{fig:bands}
\end{figure}

We consider the Rashba 2DEG in the presence of a scalar disorder potential $V(\bm{r})$, with $\bm{r} = (x,y)$, which is described by the Hamiltonian $H = H_0 + V(\bm{r})$, where
\begin{equation}
    H_0 = \frac{p^2}{2m} + \frac{\alpha}{\hbar}(\sigma_x p_y - \sigma_y p_x),
    \label{eq:H}
\end{equation}
is the clean Rashba Hamiltonian, $m$ is the effective electron mass, $\bm{p} = (p_x, p_y)$ is the momentum operator, $\alpha$ is the Rashba SOC strength, and $\bm{\sigma} = (\sigma_x, \sigma_y, \sigma_z)$ is the vector of Pauli matrices describing the spin degree of freedom. The Rashba band structure and spin texture in $k$ space are illustrated in Fig.~\ref{fig:bands}(a,b). In the numerical simulations, we discretize the Hamiltonian on a square lattice with lattice spacing $a$ and system size $L \times L$, using periodic boundary conditions. The disorder potential is modeled as a random on-site potential uniformly distributed in the range $[-W/2, W/2]$, where $W$ characterizes the disorder strength. For uncorrelated short-range disorder, we use $\expval{V(\bm{r})} = 0$ and $\langle V(\bm{r})V(\bm{r}') \rangle = \frac{1}{12}(Wa)^2 \delta(\bm{r} - \bm{r}')$, where $\langle \cdots \rangle$ denotes the ensemble average.

\begin{figure}[tb]
    \centering
    \includegraphics[width=\columnwidth]{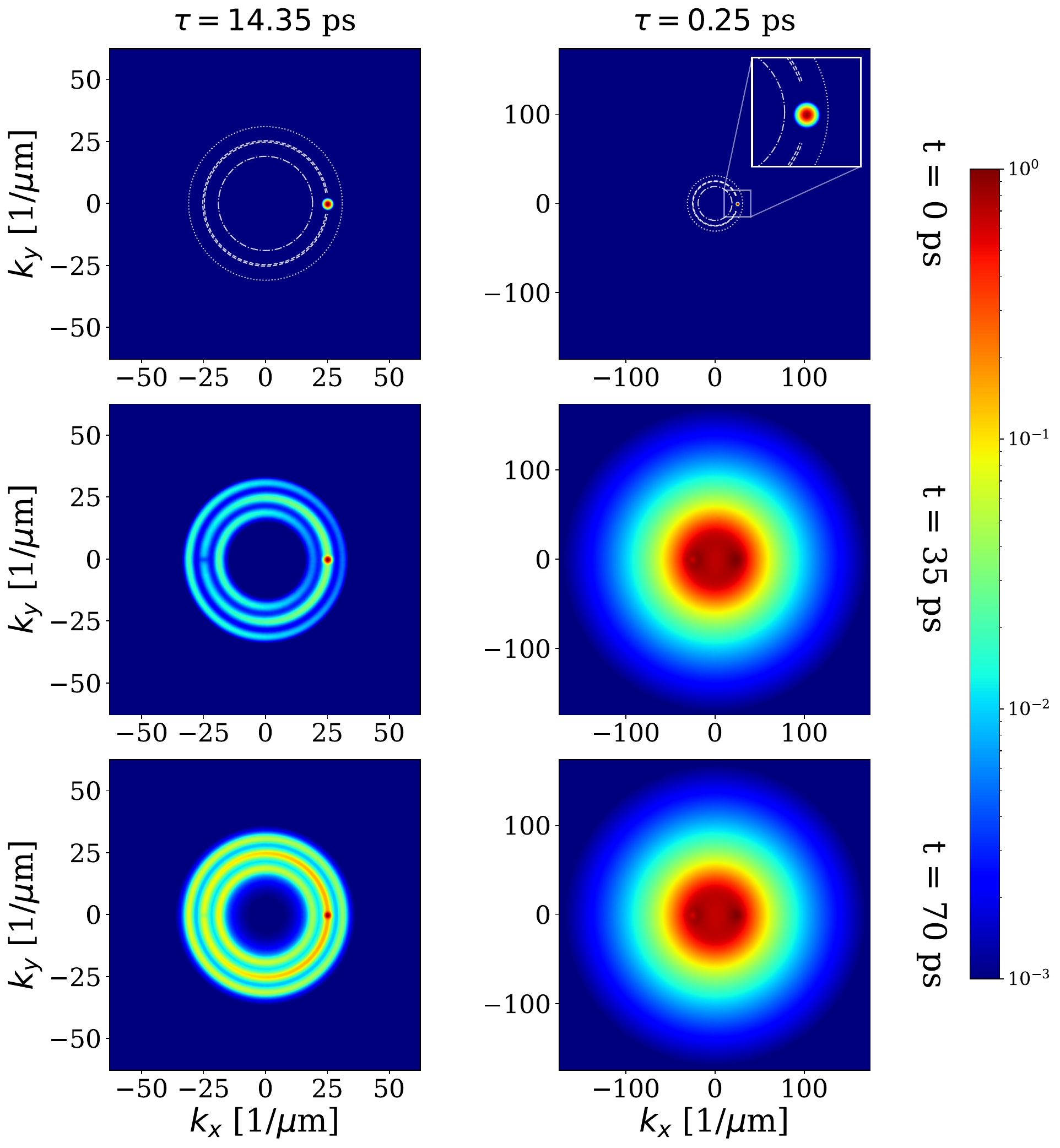}
    \caption{
        Ensemble-averaged $k$ space time evolution of $|\psi(\bm{k},t)|^2$, for illustrative times $t$ in each row. The color scale for $|\psi(\bm{k},t)|^2$ is logarithmic and normalized by the maximum value at each time. Data on the left and right panels correspond to the weak- and strong-scattering regimes, with $\tau = 14.35$~ps and $\tau = 0.25$~ps, respectively. On the top panels ($t=0$), the dashed circles indicate the constant energy contours from Fig.~\ref{fig:bands}(b). On the top right panel, the inset zooms into the initial packet.
        }
    \label{fig:densities}
\end{figure}

For the time evolution, we consider an initial Gaussian wave packet centered at the origin, with finite momentum $\bm{k}_0 = (k_0, 0)$ along the $x$ direction and spin polarized along the $z$ axis, namely,
\begin{equation}
    \psi_0(\bm{r}) = \braket{\bm{r}}{\psi_0} = \frac{e^{-r^2/4\gamma^2}}{\sqrt[4]{2\pi \gamma^2}}  e^{i \bm{k}_0 \cdot {\bm{r}}} \begin{pmatrix} 1 \\ 0 \end{pmatrix},
    \label{eq:psi0}
\end{equation}
where $\gamma$ is the real-space width of the wave packet. The initial state is illustrated in Fig.~\ref{fig:bands}(a) as a combination of opposite-helicity states at the same momentum $\bm{k}_0$. In all simulations, we use the parameters of Ref.~\cite{schliemann2007ballistic}, corresponding to InAs quantum wells with strong SOC. Specifically, $m = 0.023m_0$ ($m_0$ is the bare electron mass) and $\alpha = 10$~meVnm. The initial wave-packet width $\gamma = 1000$~nm is chosen to exceed the Rashba length $\ell_{\rm SO} = \hbar^2/m\alpha \approx 300$~nm, ensuring that the initial state is well localized in momentum space and has a well-defined initial momentum $k_0 = 25\text{~}\mu\text{m}^{-1} \gg 1/\gamma$, corresponding to energies $E = \hbar^2 k_0^2/(2m) \pm \alpha k_0 \approx 1.03 \pm 0.25$~meV, as illustrated by the constant-energy circles in Fig.~\ref{fig:bands} and Fig.~\ref{fig:densities}(a,b). The disorder strength varies in the range $0 \leq W \leq 9$~meV. The mean free path $\ell \approx v_0 \tau$ and mean free time $\tau$,
\begin{equation}
    \frac{1}{\tau} 
    = \dfrac{m W^2 a^4 n_{\rm imp}}{12 \hbar^3}
    = \dfrac{m (Wa)^2}{12 \hbar^3},
    \label{eq:tau} 
\end{equation}
calculated within the first Born approximation, provide the relevant length and time scales for the simulations. In the lattice model, the impurity density is $n_{\rm imp} = 1/a^2$, where $a$ is the lattice spacing. For the velocity, we estimate $v_0 \approx \hbar k_0/m$, neglecting SOC corrections.


For the time evolution of $\psi(\bm{r},t)$, we use a Chebyshev polynomial expansion of the time-evolution operator $U(dt) = e^{-iH dt/\hbar}$ \cite{tal-ezer1984accurate, fehske2009numerical}. This approach is efficient for large systems because it avoids full diagonalization of the Hamiltonian. Details of the numerical implementation are provided in the Appendix \ref{app:numerical}. Ensemble averaging over numerous ($N_S$) disorder samples is performed at the level of physical observables, namely position, momentum, and spin. For a generic observable $\mathcal{O}$, we write
\begin{equation}
    \expval{\mathcal{O}(t)} = \frac{1}{N_S} \sum_{i=1}^{N_S} \bra{\psi_i(t)} \mathcal{O} \ket{\psi_i(t)}.
    \label{eq:observable}
\end{equation}

Figure~\ref{fig:densities} shows the ensemble-averaged time evolution of the probability density in $k$ space, $|\psi(\bm{k},t)|^2$, for weak and strong disorder in the initial, intermediate, and long-time regimes. At $t=0$, the density is localized around the initial momentum $\bm{k}_0$. For weak disorder, the distribution remains concentrated around the constant-energy circles, whereas for strong disorder it spreads over $\bm{k}$ space. In all cases, we observe a peak-and-hole structure around $\pm \bm{k}_0$, reflecting the suppression of backscattering induced by the spin-orbit texture shown in Fig.~\ref{fig:bands}(b). Similar results were reported in Ref.~\cite{kakoi2024time}.

\begin{figure}[tb]
    \centering
    \includegraphics[width=\columnwidth]{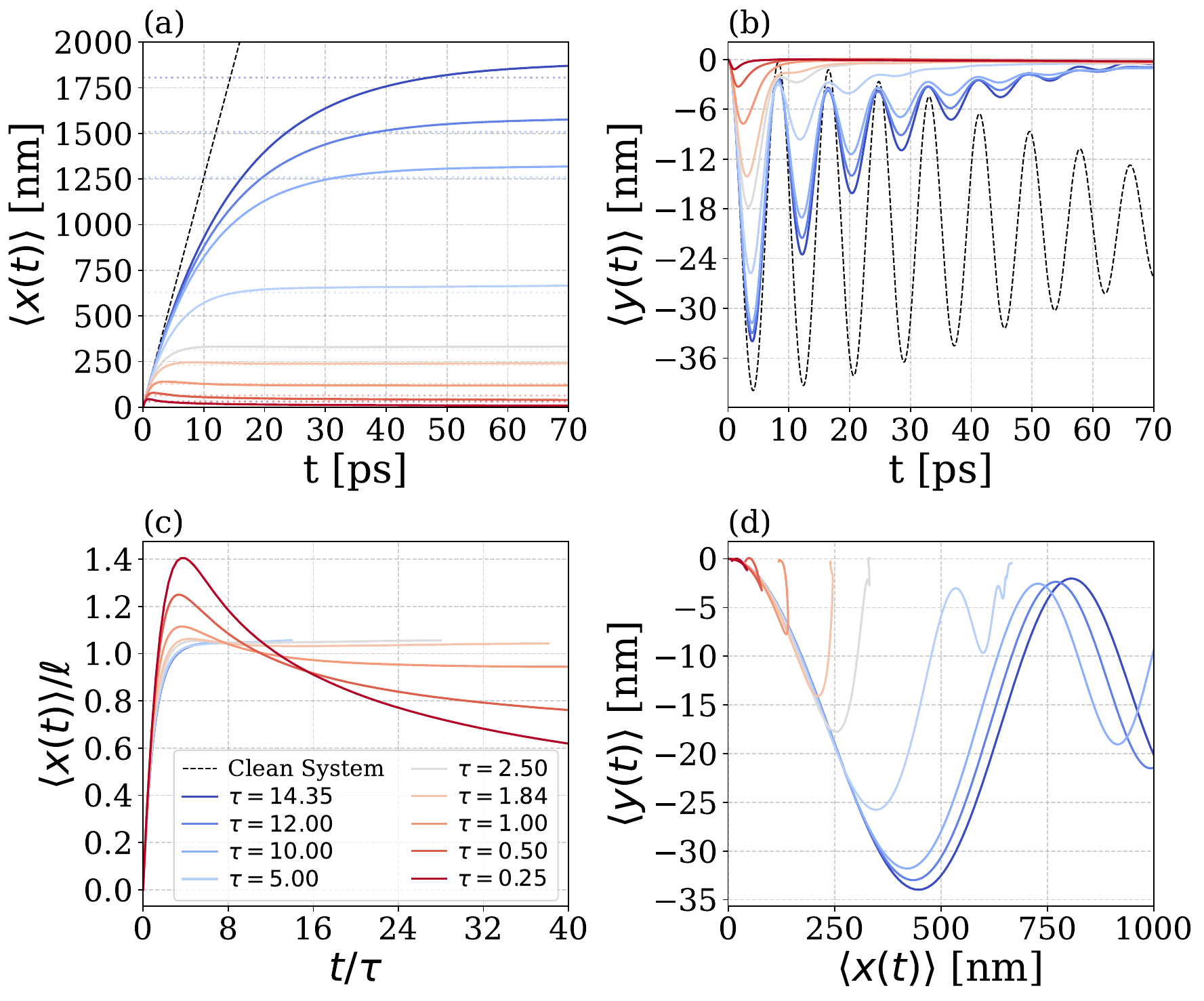}
    \caption{
        Numerical jittery quantum boomerang dynamics for ensemble-averaged trajectories at different disorder strengths $W$, color coded by their corresponding values of $\tau$. (a) Longitudinal displacement $\langle x(t) \rangle$ showing the marginal quantum boomerang effect. For weak disorder, $\langle x(t) \rangle$ shows a Drude-like relaxation approaching $\langle x(\infty) \rangle = \hbar k_0 \tau/m$ (dashed colored lines), consistent with WAL, whereas for strong disorder it shows a partial boomerang-like U-turn toward the origin, compatible with Anderson localization. (b) Transverse displacement $\langle y(t) \rangle$ showing transient Zitterbewegung that vanishes at long times $t \gg \tau$. (c) Trajectories from panel (a), with the axes scaled by the mean free path $\ell \approx v_0 \tau$ and mean free time $\tau$, emphasizing the transition from weak to strong disorder. (d) Trajectories in the $xy$ plane showing the jittery quantum boomerang effect.
        The dashed black lines on panels (a,b) correspond to clean Zitterbewegung dynamics.
        }
    \label{fig:trajectory}
\end{figure}

Figure~\ref{fig:trajectory} summarizes the disorder-averaged dynamics for increasing disorder strength, parameterized by the corresponding scattering time $\tau$. In the longitudinal direction, Fig.~\ref{fig:trajectory}(a), $\langle x(t)\rangle$ first increases and then saturates in the weak-scattering regime, giving the Drude-like response $\langle x(\infty)\rangle \approx \hbar k_0\tau/m$ derived in the next section. This weak-disorder behavior is consistent with WAL, which sustains diffusive transport. As disorder increases, the trajectories develop a partial return toward the origin, which is compatible with the marginal Anderson-localization in 2D. By contrast, in Fig.~\ref{fig:trajectory}(b), the transverse component $\langle y(t)\rangle$ shows transient Zitterbewegung at short times, followed by a return to $y=0$ for $t \gg \tau$, regardless of disorder strength.

The scaling shown in Fig.~\ref{fig:trajectory}(c), where time and displacement are expressed in units of $\tau$ and $\ell \approx v_0\tau$, reveals that the weak- and strong-disorder curves do not collapse onto a single universal trajectory. For small $\tau$, the longitudinal displacement saturates at $\langle x(t)\rangle/\ell \approx 1$, with deviations from unity arising because we estimate $\ell \approx v_0 \tau$ while neglecting SOC corrections to $v_0$. Most importantly, as $\tau$ decreases, we observe a clear transition from Drude-like saturation to boomerang-like return. Since the Rashba 2DEG belongs to the symplectic symmetry class, weak disorder is characterized by weak antilocalization (WAL), which sustains diffusive transport. The transition in $\langle x(t)\rangle$ therefore constitutes a real-time dynamical signature of the WAL-to-localization crossover: in the WAL regime the wave packet settles at a finite displacement, while increasing disorder gradually overcomes the antilocalization and drives the system toward the Anderson-localized boomerang regime. The parametric trajectories in the $xy$ plane are shown in Fig.~\ref{fig:trajectory}(d).

To identify the microscopic origin of these trajectories, next, we derive equations of motion for the weak-disorder regime, and analyze the strong-disorder regime from the perspective of Anderson localization and symmetry constraints. Throughout the discussion, we refer back to the numerical results to build a consistent picture of the dynamics across the full disorder range.

\section{Weak Disorder: \texorpdfstring{\\ }{ } Kinetic Equations of Motion}

The weak-disorder dynamics can be described using the quantum kinetic equation \cite{rammer1986quantum, rammer2007quantum, haug2008quantum, shen2014theory} within the quasi-particle approximation and the self-consistent Born approximation for disorder scattering. As shown in the Appendix \ref{app:EOM}, this framework yields the following system of equations for the center-of-mass coordinates $\expval{\bm{r}}$, momentum $\expval{\bm{k}}$, and spin polarization $\expval{\bm{\sigma}}$,
\begin{align}
    \dfrac{\partial}{\partial t}\expval{\bm{r}} &= \frac{\hbar}{m}\expval{\bm{k}} - \frac{\alpha}{\hbar} \expval{\bm{\sigma}} \times \hat{z} &&\equiv \expval{\bm{v}},
    \label{eq:velocity}
    \\
    \dfrac{\partial}{\partial t}\expval{\bm{k}} 
    &= -\dfrac{1}{\tau} \left(
    \expval{\bm{k}} - \frac{m\alpha}{\hbar^2} \expval{\bm{\sigma}} \times \hat{z}\right)
    &&\equiv -\dfrac{m}{\hbar \tau}\expval{\bm{v}},
    \label{eq:momentum}
    \\
    \dfrac{\partial}{\partial t}\expval{\bm{\sigma}} &= \expval{\bm{\Omega}(\bm{k}) \times \bm{\sigma}},
    \label{eq:spin}
\end{align}
where $\bm{\Omega}(\bm{k}) = \frac{2\alpha}{\hbar}(k_y, -k_x, 0)$ is the effective magnetic field due to the Rashba SOC. These EOMs do not form a closed set, as the spin equation contains mixed spin-momentum averages of the form $\langle k_\mu \sigma_\nu \rangle$. Nevertheless, they are useful for qualitative and asymptotic analysis.

\begin{figure}[tb]
    \centering
    \includegraphics[width=\columnwidth]{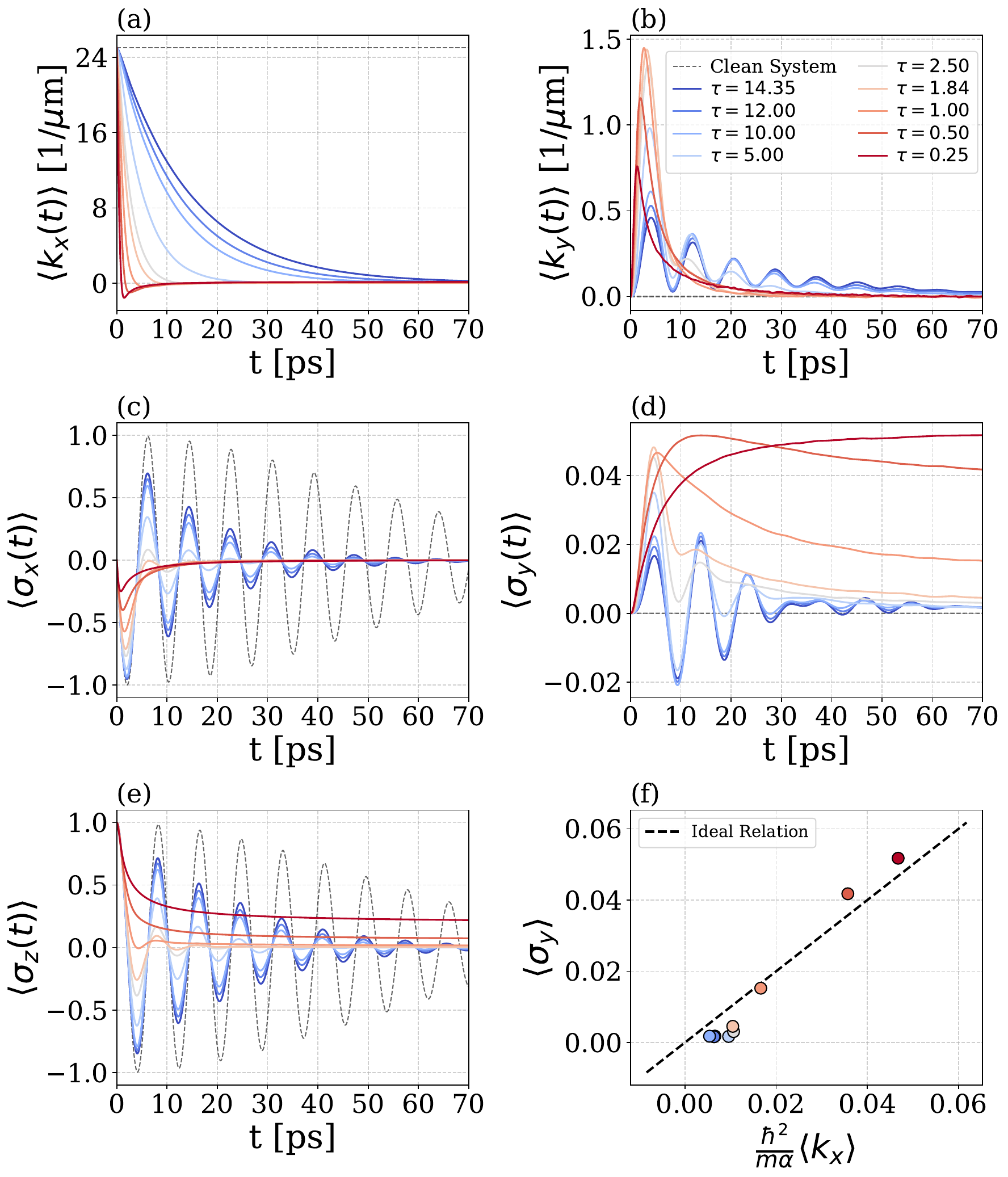}
    \caption{
        (a,b) Ensemble-averaged dynamics of momentum $\langle k_x(t) \rangle$ and $\langle k_y(t) \rangle$, and (c-e) spin $\langle \sigma_x(t) \rangle$, $\langle \sigma_y(t) \rangle$, and $\langle \sigma_z(t) \rangle$. Data for different disorder strengths $W$ are color coded and identified by their values of $\tau$. The transition from weak to strong disorder is seen in the disappearance of oscillations in $\langle k_y(t) \rangle$ and in the negative dip of $\langle k_x(t) \rangle$, which leads to the partial quantum boomerang effect observed in $\langle x(t) \rangle$ in Fig.~\ref{fig:trajectory}. The spin dynamics show similar behavior, with oscillations vanishing for strong disorder. Both $\langle k_x(t) \rangle$ and $\langle k_y(t) \rangle$ show exponential relaxation at long times as disorder scattering isotropizes the momentum distribution. However, only $\langle k_y(t) \rangle$ reaches zero asymptotically, while $\frac{\hbar^2}{m\alpha}\langle k_x \rangle = \langle \sigma_y \rangle$, as shown in panel (f), where the dashed line denotes the ideal asymptotic relation and the colored circles show numerical data for different disorder strengths. Small deviations from the dashed line are expected because convergence at long $t$ is slow and not fully reached in the simulations.
        }
    \label{fig:k_sigma}
\end{figure}

In the absence of disorder ($1/\tau \rightarrow 0$), the EOMs for the spin-dependent velocity and the spin precession lead to the Zitterbewegung effect, which is shown as dashed black curves in the trajectories of Fig.~\ref{fig:trajectory}(a,b) and the momentum and spin dynamics in Fig.~\ref{fig:k_sigma}. While the clean system symmetries dictate that $\langle k_y(t) \rangle \equiv 0$ and $\langle \sigma_y(t) \rangle \equiv 0$ for all $t$, the remaining non-zero components exhibit a transient, rather than perpetually oscillating behavior. This decay occurs because the two helicity branches of the Rashba band structure have distinct velocity-dependent dispersion relations, making the wave-packet components split in real space due to their different group velocities, thus, suppressing the interference that sustains the Zitterbewegung oscillations. Namely, the transverse displacement $\langle y(t) \rangle$ in Fig.~\ref{fig:trajectory}(b) levels off to a finite, ballistic side-jump plateau, whereas the corresponding spin expectations $\langle \sigma_x(t) \rangle$ and $\langle \sigma_z(t) \rangle$ in Fig.~\ref{fig:k_sigma}(c,e) exhibit damped oscillations that gradually vanish even in the absence of impurity scattering.

For finite disorder, the finite values of $\langle k_y(t) \rangle$ and $\langle \sigma_y(t) \rangle$ emerge due to SOC-induced anisotropic scattering \cite{kakoi2024time}. For weak disorder, the EOM for $\expval{\bm{k}}$ takes the form of a viscous damping, which suppresses the Zitterbewegung and forces the particle to return to the origin along $y$ at long times. This can be seen by combining Eqs.~\eqref{eq:velocity} and \eqref{eq:momentum} and integrating over time, which yields $\expval{\bm{r}(\infty)} = (\hbar\tau/m)[\bm{k}_0 - \expval{\bm{k}(\infty)}]$. Then, the stationary conditions, $\partial_t \expval{\bm{r}} = 0$ and $\partial_t \expval{\bm{k}} = 0$, imply the asymptotic constraint $\expval{\bm{k}(\infty)} = (m\alpha/\hbar^2) \expval{\bm{\sigma}(\infty)} \times \hat{z}$. For our parameter set, and taking the extreme value $|\expval{\bm{\sigma}(\infty)}| = 1$, it gives $|\expval{\bm{k}(\infty)}| \lesssim 3 \times 10^{-3}$~nm$^{-1} \ll k_0$. This allows us to approximate
\begin{align}
    \expval{x(\infty)} &\approx \frac{\hbar\tau}{m} k_0,
    \label{eq:x_infty} 
    \\
    \expval{y(\infty)} &= -\dfrac{\hbar\tau}{m} \expval{k_y(\infty)}
    \approx \dfrac{\alpha \tau}{\hbar} \langle \sigma_x(\infty) \rangle \approx 0.
    \label{eq:y_infty}
\end{align}
Thus, along $x$ the dynamics saturate at finite values, as shown by the dashed lines in Fig.~\ref{fig:trajectory}(a). By contrast, the vanishing of $\expval{y(\infty)}$ reflects the disorder-induced isotropization of the momentum distribution at long times, which leads to $\langle k_y(\infty) \rangle \approx 0$ and, due to the asymptotic constraint above, $\langle \sigma_x(\infty) \rangle \approx 0$. Moreover, for strong disorder, as shown in the next section, symmetry constraints and Anderson localization identically enforce both $\langle k_y(\infty) \rangle \equiv 0$ and $\langle \sigma_x(\infty) \rangle \equiv 0$.

Figure~\ref{fig:k_sigma} shows that, in both momentum and spin dynamics, the crossover from weak to strong disorder appears as a transition from oscillatory motion to exponential relaxation. Moreover, for strong disorder, $\langle k_x(t) \rangle$ develops a negative dip, which produces the boomerang-like reversal in $\langle x(t) \rangle$. Interestingly, $\langle \sigma_y(\infty) \rangle$ and $\langle \sigma_z(\infty) \rangle$ saturate at finite values for strong disorder. This occurs when the spin does not have enough time to precess between scattering events, namely for small $\tau \ll \tau_{\rm soc} = \pi \hbar/\alpha k_0 \approx 8$~ps. This transition matches the crossover from D'yakonov-Perel to Elliott-Yafet spin-relaxation mechanisms reported in Ref.~\cite{kakoi2024time}.

\section{Strong disorder: localization and symmetry constraints}

As the disorder strength $W$ increases, Anderson localization gradually sets in and the EOMs above slowly break down. To analyze this regime in the asymptotic limit $t\rightarrow \infty$, we extend the symmetry and localization arguments of the quantum boomerang effect in Refs.~\cite{prat2019quantum, noronha2022ubiquity} to include SOC.

To emphasize the initial state parameters for the longitudinal momentum $k_0$ and spin $s_z$, in this section, when necessary, we explicitly add $(k_0, s_z)$ to the arguments of the disorder-averaged observables (position, momentum and spin) from Eq.~\eqref{eq:observable}. Namely, $\expval{\mathcal{O}(k_0, s_z, t)} \equiv \expval{\mathcal{O}(t)}$.

\begin{table}[tb]
\caption{Transformation of the position, momentum and spin observables, under the $\mathcal{M}_x \mathcal{T}$ symmetry.}
\begin{ruledtabular}
\begin{tabular}{c|ccccccc}
 Observable & $\phantom{+}x$ & $\phantom{+}y$ & $\phantom{+}k_x$ & $\phantom{+}k_y$ & $\phantom{+}\sigma_x$ & $\phantom{+}\sigma_y$ & $\phantom{+}\sigma_z$ \\
\hline
$\mathcal{M}_x\mathcal{T}$ & $-x$ & $\phantom{+}y$ & $\phantom{+}k_x$ & $-k_y$ & $-\sigma_x$ & $\phantom{+}\sigma_y$ & $\phantom{+}\sigma_z$ \\
\end{tabular}
\end{ruledtabular}
\label{tab:symmetries}
\end{table}

We begin with the symmetry constraints. In the absence of disorder, the Rashba 2DEG model from Eq.~\eqref{eq:H} is invariant under mirror $\mathcal{M}_x$ and time-reversal $\mathcal{T}$ symmetries. Since the disorder potential is scalar, it preserves both $\mathcal{M}_x$ and $\mathcal{T}$ on average. Therefore, the disorder-averaged dynamics must be covariant under these symmetries, which imposes specific transformation rules for the observables (position, momentum, and spin). In particular, the composite symmetry $\mathcal{M}_x \mathcal{T}$ implies that the disorder-averaged dynamics must satisfy 
\begin{equation}
    \expval{\mathcal{O}(k_0, s_z, t)} = \pm \expval{\mathcal{O}(k_0, s_z,-t)},
    \label{eq:symmetry_constraint}
\end{equation}
where the sign depends on the parity of the observable $\mathcal{O}$ under $\mathcal{M}_x \mathcal{T}$, as summarized in Table~\ref{tab:symmetries}. These conditions are exact for all times and disorder strengths, and they are independent of the presence of Anderson localization.

Now, considering Anderson localization, the dynamics for $\expval{\mathcal{O}(k_0, s_z, t)}$ can be expanded as \cite{prat2019quantum, noronha2022ubiquity}
\begin{align}
    \expval{\mathcal{O}(k_0, s_z, t)} &= \sum_{m,n} \overline{c_{m}^* c_n e^{-\frac{i}{\hbar}(\varepsilon_m - \varepsilon_n)t} \bra{\phi_m} \mathcal{O} \ket{\phi_n}},
    \\
    \expval{\mathcal{O}(k_0, s_z, \infty)} &\approx \sum_{n} \overline{|c_n|^2 \bra{\phi_n} \mathcal{O} \ket{\phi_n}},
    \label{eq:O_dephasing}
\end{align}
where the initial conditions $(k_0, s_z)$ are implicit in the coefficients $c_n$. Here, the wave packet is expanded in the eigenbasis of the disordered Hamiltonian as $\ket{\psi_0} = \sum_n c_n \ket{\phi_n}$, with $H\ket{\phi_n} = \varepsilon_n \ket{\phi_n}$. Assuming localized eigenstates, the phase factors $e^{-\frac{i}{\hbar}(\varepsilon_m-\varepsilon_n)t}$ oscillate with finite frequencies for $m\neq n$. After long-time and disorder averaging, these oscillations cancel the off-diagonal contributions, yielding Eq.~\eqref{eq:O_dephasing}. Since Eq.~\eqref{eq:O_dephasing} is the same for forward and backward in time, it implies the asymptotic ($t\rightarrow \pm \infty$) constraint
\begin{align}
    \expval{\mathcal{O}(k_0, s_z, \infty)} &= \expval{\mathcal{O}(k_0, s_z, -\infty)}.
    \label{eq:AL_constraint}
\end{align}
Note that this constraint relies on Anderson localization, which is marginal in 2D. Therefore, it is strictly valid only for $W\rightarrow \infty$, although signatures of its effect are expected to emerge gradually as $W$ increases.

In the asymptotic limit, $t \rightarrow \infty$, the symmetry and localization constraints from Eqs.~\eqref{eq:symmetry_constraint} and \eqref{eq:AL_constraint} must be simultaneously satisfied. This condition is violated for observables that exhibit odd parity under $\mathcal{M}_x \mathcal{T}$ (see Table~\ref{tab:symmetries}), unless it identically vanishes at $t\rightarrow\infty$. Therefore, these constraints impose that $\expval{x(k_0, s_z, \infty)} = 0$, $\expval{k_y(k_0, s_z, \infty)} = 0$, and $\expval{\sigma_x(k_0, s_z, \infty)} = 0$. Particularly, the vanishing of $\expval{x(k_0, s_z,\infty)}$ is the hallmark of the longitudinal quantum boomerang effect, which is only partial in the marginally localized 2D case and reaches the perfect return only in the limit of infinite disorder strength, as seen in Fig.~\ref{fig:trajectory}(c).

In contrast, the even parity of $y$, $k_x$, $\sigma_y$, and $\sigma_z$ under $\mathcal{M}_x \mathcal{T}$ does not yield a direct constraint on their asymptotic values. However, the weak-disorder EOMs show that $\expval{y(k_0, s_z, \infty)}$ is proportional to $\expval{k_y(k_0, s_z, \infty)}$ [or $\expval{\sigma_x(k_0, s_z, \infty)}$, see Eq.~\eqref{eq:y_infty}], and both are constrained to vanish by symmetry and localization. Assuming this proportionality extends approximately into the strong-disorder regime, it follows that $\expval{y(k_0, s_z, \infty)} = 0$, yielding the transverse quantum boomerang effect. Indeed, the numerical data in Fig.~\ref{fig:k_sigma}(f) show that the asymptotic relation $\frac{\hbar^2}{m\alpha} \expval{k_x(\infty)} = \expval{\sigma_y(\infty)}$, derived from the stationary EOM conditions for weak disorder, remains approximately valid for strong disorder.

\section{Conclusions}

In summary, we have studied the real-time evolution of a spin-polarized wave packet in a disordered Rashba 2DEG and identified the \textit{jittery quantum boomerang effect}. We show that the longitudinal and transverse quantum boomerang dynamics follow distinct mechanisms. Along $x$, the wave packet crosses over from the weak-disorder Drude-like regime — characteristic of weak antilocalization (WAL) in the symplectic class  — to a boomerang-like return as disorder increases. This crossover in $\langle x(t)\rangle$ provides a real-time dynamical signature of the WAL-to-localization transition, consistent with the onset of marginal Anderson localization in two dimensions. Along $y$, clean-system Zitterbewegung is progressively damped by impurity scattering and vanishes at long times, thereby suppressing the ballistic side jump for arbitrary disorder strength.

We find that our jittery quantum boomerang dynamics qualitatively match the experimental observations of Ref.~\cite{werake2011observation}, where transient motion along $y$ was observed at short times $t < \tau$ and interpreted as the \textit{intrinsic spin Hall effect} in the time domain. The striking similarity between those experimental trajectories and our numerical results suggests that our model captures the essential physics of the experiment, despite the absence of a well-defined Fermi surface in our single-particle description.

We have not considered external electric fields $\bm{F}$, which would continuously shift the momentum distribution and require energy-relaxation mechanisms, such as phonon scattering, to reach a steady state. However, the vanishing of the transverse Zitterbewegung for weak disorder is expected to remain valid even in the presence of an electric field $\bm{F} \parallel \hat{x}$, since it would not affect the EOMs for $\expval{y}$ and $\expval{k_y}$ that govern the transverse dynamics. 

\begin{acknowledgments}
    We acknowledge useful discussions with Makoto \mbox{Kohda}.
    This work was supported by the Brazilian funding agencies FAPEMIG, CNPq, CAPES.
\end{acknowledgments}

\bibliography{main, refArxiv}


\appendix
\setcounter{equation}{0}
\renewcommand{\theequation}{A\arabic{equation}}

\section{Numerical implementation}
\label{app:numerical}

The time-evolution operator $U(dt) = e^{-iH dt/\hbar}$ is implemented using a Chebyshev polynomial expansion \cite{tal-ezer1984accurate, fehske2009numerical}, whose recurrence relation yields the following recursive expression for the time-evolved state:
\begin{align}
    \ket{\psi(t+dt)} &= J_0(\delta)\ket{\phi_0} + 2\sum_{n=1}^{N \gg \delta} (-i)^n J_n(\delta) \ket{\phi_n},
    \label{eq:psit}
    \\
    \ket{\phi_n} &= 2\tilde{H} \ket{\phi_{n-1}} - \ket{\phi_{n-2}},
    \label{eq:phin}
\end{align}
with $\ket{\phi_1} = \tilde{H}\ket{\phi_0}$ and $\ket{\phi_0} = \ket{\psi(t)}$. Here, $J_n$ is the Bessel function of the first kind, $\delta = \Delta E dt/\hbar$ and $\tilde{H} = (H - E_{\rm C})/\Delta E$ are the rescaled time and Hamiltonian, respectively, with $E_{\rm C} = (E_{\rm max} + E_{\rm min})/2$ and $\Delta E = (E_{\rm max} - E_{\rm min})/2$ estimated from the energy spectrum of $H$. The sum is truncated at a finite order $N \gg \delta$ to ensure convergence, since $J_n(\delta)$ decays rapidly for $n > \delta$. 

For different values of the disorder strength $W$, we adjust the system size $L$ and lattice spacing $a$ to ensure that the wave packet does not interact with the boundaries in either real or reciprocal space during the time evolution. Overall, these parameters remain within the ranges $8000 \leq L \leq 30000$~nm and $18 \leq a \leq 60$~nm. The ensemble averaging is performed with a sufficiently large number of disorder samples $N_S$ to ensure convergence of the observables, with typical values of $N_S$ ranging from $5000$ to $30000$. We have verified that the results are independent of the specific choice of parameters within these ranges.

\section{Equations of motion}
\label{app:EOM}

Here, we provide the technical derivation of the equations of motion for the spin-polarized wave packet in the weak-disorder regime. We use the quantum kinetic equation \cite{rammer1986quantum, rammer2007quantum, haug2008quantum, shen2014theory} for the spinful Wigner distribution function $f(\bm{q}, \bm{k}, t) = \sum_i f_i(\bm{q}, \bm{k}, t) \sigma_i$, where $i = (0, x, y, z)$, $\sigma_0$ is the identity matrix, and $\bm{q}$ is the Fourier conjugate of the center-of-mass coordinate $\bm{r}$.

The kinetic equation is governed by
\begin{align}
    \frac{\partial}{\partial t} f -\frac{1}{i\hbar}[H_0, f] + \frac{1}{2\hbar}\{\nabla_k H_0, \nabla_r f\} = I[f],
\\
    I[f] = \frac{i}{4\pi\hbar} \int dE\Big[ \{A, \Sigma^K\} - \{\Gamma, G^K\} \Big],
\end{align}
where $I[f]$ is the collision integral, and $A$ and $\Gamma$ are the spectral function and scattering rate, respectively. Above, we have neglected the external electric field term for simplicity. Applying the quasiparticle and self-consistent Born approximations, we define the spectral function $A$ and self-energy $\Sigma^\nu$ as
\begin{align}
    A &= 2\pi \exp\Big[-H_{\rm soc}\frac{\partial}{\partial E}\Big] \delta(E - \varepsilon_0(k)),
\\
    \Sigma^\nu &= \frac{n_{\rm imp} (Wa)^2}{12 L^2} \sum_{\bm{k}} G^\nu,
\end{align}
where $\varepsilon_0(k) = \hbar^2 k^2/2m$. In terms of the vector $\bm{f} = (f_0, f_x, f_y, f_z)^T$, the kinetic equation takes the compact form \cite{deassis2021spin}
\begin{multline}
    \Big[\frac{\partial}{\partial t} + i \frac{\hbar}{m} \bm{q} \cdot \bm{k} + \mathcal{K} \Big]\bm{f}(\bm{q}, \bm{k}, t) = 
\\
    -\frac{1}{\tau} \Big[\bm{f}(\bm{q}, \bm{k}, t) - (1+\mathcal{C})\expval{\bm{f}(\bm{q}, k, t)}_\theta\Big],
\end{multline}
where $\expval{\dots}_\theta$ denotes the angular average in $k$-space, and the Rashba SOC and collision correction matrices are defined as
\begin{align}
    \mathcal{K} &= \frac{\alpha}{\hbar} \begin{pmatrix} 0 & iq_{y} & -iq_{x} & 0\\ iq_{y} & 0 & 0 & 2k_{x}\\ -iq_{x} & 0 & 0 & 2k_{y}\\ 0 & -2k_{x} & -2k_{y} & 0 \end{pmatrix},
\\
    \mathcal{C} &= \begin{pmatrix} 0 & \alpha k_{y} & -\alpha k_{x} & 0\\ \alpha k_{y} & 0 & 0 & 0\\ -\alpha k_{x} & 0 & 0 & 0\\ 0 & 0 & 0 & 0 \end{pmatrix} \frac{\partial}{\partial\varepsilon(k)}.
\end{align}

The mean values of position, momentum, and spin are expressed as integrals over reciprocal space
\begin{align}
    \expval{\bm{r}(t)} &= \int i\nabla_{q} f_{0}(\bm{q},\bm{k},t)d^{2}qd^{2}k,
    \\
    \expval{\bm{k}(t)} &= \int \bm{k} f_{0}(\bm{q},\bm{k},t)d^{2}qd^{2}k,
    \\
    \expval{\sigma_{\nu}(t)} &= \int f_{\nu}(\bm{q},\bm{k},t)d^{2}qd^{2}k,
\end{align}
with the normalization condition $\int f_{0}(\bm{q},\bm{k},t)d^{2}qd^{2}k = 1$. 

To obtain the EOMs for the momentum, we consider the scalar component ($f_0$) of the kinetic equation in the large-wave-packet limit ($\bm{q} \rightarrow 0$),
\begin{equation}
    \partial_t f_0 = -\frac{1}{\tau} \Big[f_0 - \expval{f_0}_\theta 
    - \alpha k_{y}\frac{\partial \langle f_{x}\rangle_{\theta}}{\partial\varepsilon(k)} + \alpha k_{x}\frac{\partial \langle f_{y}\rangle_{\theta}}{\partial\varepsilon(k)}
    \Big]
\end{equation}
Multiplying by $k_\nu$ (for $\nu = x, y$) and integrating over the full $(\bm{q},\bm{k})$-space ($d^2q \, d^2k = d^2q\, k \, dk \, d\theta$), the angular average yields $\langle k_{x}k_{y}\rangle_{\theta}=0$, leading to
\begin{align}
    \partial_{t}\langle k_{x}\rangle &= -\frac{1}{\tau}\Big[\langle k_{x}\rangle + \alpha \int k_x^2 \frac{\partial \langle f_{y}\rangle_{\theta}}{\partial\varepsilon(k)} d^2k \Big], 
\\
    \partial_{t}\langle k_{y}\rangle &= -\frac{1}{\tau}\Big[\langle k_{y}\rangle - \alpha \int k_y^2 \frac{\partial \langle f_{x}\rangle_{\theta}}{\partial\varepsilon(k)} d^2k \Big].
\end{align}

To evaluate the integral on the right-hand side, we first perform the angular average in cylindrical coordinates. For the $k_x$-component we get
\begin{equation}
    \alpha \int k_{x}^{2}\frac{\partial \langle f_{y}\rangle_{\theta}}{\partial\varepsilon(k)} d^2k = \alpha  \pi \int_0^\infty k^{3} \frac{\partial \langle f_{y}\rangle_{\theta}}{\partial\varepsilon(k)} dk.
\end{equation}
Then, we change variables from $\varepsilon$ to $k$ in the derivative considering the dispersion relation $\varepsilon(k) = \frac{\hbar^{2}}{2m}k^{2} + \eta\alpha k$. Keeping only the lowest order in $\alpha$, we approximate $\varepsilon(k) \approx \frac{\hbar^{2}}{2m}k^{2}$. Using the chain rule $\frac{\partial}{\partial\varepsilon} = \frac{\partial k}{\partial \varepsilon} \frac{\partial}{\partial k}$, we get
\begin{equation}
    \alpha \pi \int_0^\infty k^{3} \frac{\partial \langle f_{y}\rangle_{\theta}}{\partial \varepsilon} dk = \pi \frac{m}{\hbar^2} \int_0^\infty k^{2} \frac{\partial \langle f_{y}\rangle_{\theta}}{\partial k} dk.
\end{equation}
Applying integration by parts and assuming the boundary terms vanish yields
\begin{multline}
    -\alpha \pi \frac{m}{\hbar^{2}} \int_0^\infty 2k \langle f_{y}\rangle_{\theta} dk = -\frac{m\alpha}{\hbar^{2}} \left( 2\pi \int_0^\infty \langle f_{y}\rangle_{\theta} k \, dk \right) 
\\
    = -\frac{m\alpha}{\hbar^{2}}\langle\sigma_{y}\rangle.
\end{multline}
Similarly, the $y$-component evaluates to
\begin{equation}
    \alpha \int k_{y}^{2}\frac{\partial\langle f_{x}\rangle_{\theta}}{\partial\varepsilon(k)} d^2k = -\frac{m\alpha}{\hbar^{2}}\langle\sigma_{x}\rangle.
\end{equation}
Collecting the results, we obtain the EOM for the momentum in Eq.~\eqref{eq:momentum} of the main text,
\begin{equation}
    \dfrac{\partial}{\partial t}\expval{\bm{k}} 
        = -\dfrac{1}{\tau} \left(
        \expval{\bm{k}} - \frac{m\alpha}{\hbar^2} \expval{\bm{\sigma}} \times \hat{z}\right),
\end{equation}

To derive the spatial EOM for $\bm{r}$, we consider again the scalar component ($f_0$) of the kinetic equation. However, the gradient operator $\bm{r} \rightarrow i\nabla_{\bm{q}}$ must now be applied prior to taking the large-wave-packet limit ($\bm{q} \rightarrow 0$). Isolating the time derivative for $f_0$ from the compact kinetic equation yields
\begin{multline}
    \frac{\partial f_0}{\partial t} = -i \frac{\hbar}{m} (\bm{q} \cdot \bm{k}) f_0 - i\frac{\alpha}{\hbar} (q_y f_x - q_x f_y)
\\
    -\frac{1}{\tau} \Big[f_0 - \expval{f_0}_\theta 
    - \alpha k_{y}\frac{\partial \langle f_{x}\rangle_{\theta}}{\partial\varepsilon(k)} + \alpha k_{x}\frac{\partial \langle f_{y}\rangle_{\theta}}{\partial\varepsilon(k)}
    \Big].
\end{multline}
Applying the integration $\int d^{2}q d^{2}k \, i\nabla_{\bm{q}}$ to both sides, the collision integral terms between square brackets vanish due to angular average in k-space. On the remaining terms, the action of the $i\nabla_{\bm{q}}$ operator yields
\begin{multline}
    i\nabla_{\bm{q}} \left[ -i \frac{\hbar}{m} (\bm{q} \cdot \bm{k}) f_0 - i\frac{\alpha}{\hbar} (q_y f_x - q_x f_y) \right] 
\\
    = \frac{\hbar}{m} \bm{k} f_0 - \frac{\alpha}{\hbar} (\bm{f} \times \hat{z}),
\end{multline}
Integrating over $(\bm{q}, \bm{k})$ results in Eq.~\eqref{eq:velocity} of the main text,
\begin{equation}
    \frac{\partial}{\partial t}\expval{\bm{r}} = \frac{\hbar}{m}\expval{\bm{k}} - \frac{\alpha}{\hbar} \expval{\bm{\sigma}} \times \hat{z}.
\end{equation}

Similarly, to evaluate the spin dynamics, we project the kinetic equation onto the Pauli vector components $\bm{f}_\sigma = (f_x, f_y, f_z)^T$. Because there is no spatial gradient involved in the spin definition, we can directly employ the large-wave-packet limit ($\bm{q} \rightarrow 0$). The terms from the collision integral vanish in the k-space angular average, as they do for the EOM for $\bm{r}$. Therefore, the time evolution is governed by the lower $3 \times 3$ block of the matrix $\mathcal{K}$. By defining the momentum-dependent Rashba precession vector $\bm{\Omega}(\bm{k}) = \frac{2\alpha}{\hbar} (k_y, -k_x, 0)$, the action of $\mathcal{K}$ on $\bm{f}_\sigma$ is isomorphic to a vector cross product
\begin{equation}
    -\mathcal{K}_{3\times 3} \bm{f}_\sigma = \bm{\Omega}(\bm{k}) \times \bm{f}_\sigma.
\end{equation}
Substituting this geometric identity into the kinetic equation yields
\begin{equation}
    \frac{\partial \bm{f}_\sigma}{\partial t} = \bm{\Omega}(\bm{k}) \times \bm{f}_\sigma.
\end{equation}
Integrating over ($\bm{q}, \bm{k})$ recovers the pure precession dynamics for the average spin polarization, given by Eq.~\eqref{eq:spin} of the main text,
\begin{equation}
    \frac{\partial}{\partial t}\expval{\bm{\sigma}} = \expval{\bm{\Omega}(\bm{k}) \times \bm{\sigma}}.
\end{equation}

\end{document}